# SUPPORTING BELONGING IN SOFTWARE ENGINEERING THROUGH ROLE MODELS EXPOSURE

*Ronnie de Souza Santos*
University of Calgary
ronnie.desouzasantos@ucalgary.ca

**Abstract** – *Role models are widely discussed in educational research as influential in students' identity development and sense of belonging, yet less attention has been given to how role model visibility can be systematically embedded within everyday engineering instruction. This paper presents an analytic autoethnographic account of integrating historically grounded role models into routine software engineering teaching practice. Drawing on reflective memos and instructional artifacts across multiple course offerings, we characterize how brief, topic-aligned contextualizations of pioneers were incorporated into core technical lectures without altering learning objectives or assessments. The findings indicate that this structurally embedded approach functioned as a low-disruption pedagogical practice that aligned representation with disciplinary substance, situating diverse contributors as foundational to the development of software architecture. The integration was iterative and refined across semesters to strengthen topic alignment and instructional flow. These results suggest that embedding historically grounded representation within technical content may serve as a practical mechanism for supporting inclusivity while preserving technical rigor in engineering education.*

**Paper Type**: Educational Practice

## 1. INTRODUCTION

Role models have been widely investigated in educational research as influential figures in students' identity formation, academic motivation, and long-term achievement [6, 10, 16]. The literature shows that exposure to role models is associated with stronger achievement-oriented goals, greater engagement in academic activities, and improved academic performance over time [16]. In engineering education, structured exposure to peer role models has been linked to higher course grades and increased enrollment in subsequent engineering courses, with particularly notable benefits for women and students from underrepresented backgrounds [10]. In this sense, role models communicate what is attainable for members of specific social groups and contribute to how students imagine their place within academic and professional domains [10, 13, 16].

Within software engineering, concerns about representation and persistence have motivated the implementation of mentoring and role model programs aimed at improving recruitment, retention, and community building [14]. Characterizations of female students in software engineering highlight recurring patterns, including high academic achievement, limited prior programming exposure, lower self-confidence despite strong competence, and experiences of stereotype-related pressure [1, 8]. These findings suggest that academic performance alone does not guarantee a strong sense of belonging and that structural and social factors shape students' perceptions of their position within the field [8, 14].

Beyond educational contexts, role models in software engineering practice are valued not only for technical expertise but also for moral values, professional integrity, and human qualities [11]. This broader understanding of role modeling indicates that representation operates at multiple levels, influencing not only skill development but also identity construction and perceptions of legitimacy within the profession [11]. These studies indicate that representation and identity function across both educational and professional environments, shaping perceptions of belonging in software engineering [10, 11, 16].

Despite the growing body of work on mentoring programs, peer initiatives, and role model interventions in engineering and software education [10, 12, 14], less attention has been given to how role models can be systematically embedded into everyday software engineering classes and explicitly connected to technical course content. Therefore, in this paper, we adopt an autoethnographic approach to reflect on our experience incorporating diverse role models into daily software engineering teaching practice and associating them with technical aspects of the course, with the goal of describing how such practices may contribute to strengthening the sense of belonging of students from underrepresented groups in the field.

## 2. LITERATURE REVIEW

Role models in education are conceptualized as socially visible individuals whose presence provides information about opportunity structures, professional

Peer reviewed

trajectories, and normative expectations within a field [16]. Rather than functioning solely as sources of inspiration, role models contribute to how students interpret the alignment between their social identity and specific academic or professional positions [16]. Empirical research in higher education demonstrates that demographic representation among faculty can influence students' academic choices and field engagement, indicating that visibility and similarity shape educational pathways [3]. In engineering education, role models are further described as influencing expectancy beliefs and perceived attainability of disciplinary goals, operating through social comparison mechanisms and perceived similarity [6, 10]. In engineering contexts, role modeling has been operationalized through structured mentoring systems, peer-led initiatives, and institutional programs designed to increase visibility and community integration [14].

In software engineering education, role modeling has been reported as a strategy to support students from underrepresented groups in the field, including women, Black individuals, and LGBTQIA+ students, by strengthening networks of peer interaction and increasing exposure to relatable figures within the discipline [8, 12, 14]. The reported benefits of role model exposure extend beyond individual motivation to include shifts in future orientation, perceived legitimacy within a field, and interpretations of attainable professional trajectories [16]. In professional software engineering contexts, for example, role models are described as shaping ethical standards, professional norms, and value systems, suggesting that their influence encompasses identity construction and normative alignment in addition to technical development [6, 11]. However, the literature also documents limitations. Increased visibility does not automatically eliminate stereotype pressures or disparities in self-evaluation among software engineering students [1, 8]. Faculty representation effects are shown to vary depending on institutional context and implementation conditions [3]. Diversity analyses caution that without addressing broader structural inequalities, role modeling initiatives may have constrained or uneven impact across different groups [12, 13]. These findings indicate that role models function as context-dependent mechanisms whose effects are mediated by institutional design, structural alignment, and the interaction between representation and lived experience [16].

## 3. METHOD

This study adopts an analytic autoethnographic approach to investigate how the systematic integration of diverse role models into routine software engineering instruction may contribute to fostering students' sense of belonging. Autoethnography is an empirical method that describes and systematically analyzes personal experience in order to understand broader cultural practices [5]. In contrast to traditional ethnography, where an external researcher observes a social setting, analytic autoethnography positions the researcher as a complete member of the setting who reflexively examines their own practice while maintaining a commitment to theoretical interpretation [2]. In this study, the instructor documents and analyzes the lived experience of embedding role model visibility into everyday technical teaching within software engineering education. Consistent with analytic expectations in autoethnographic research, the narrative is not presented solely as personal reflection but is explored through a structured conceptual lens grounded in scholarship on role models, representation, and belonging in engineering education [2, 5].

### 3.1. Context and Setting

The study was conducted in two educational settings at a Canadian research-intensive university: a third-year undergraduate course on Software Architecture and a course-based Master-level Software Engineering course. The initiative began in Winter 2024 and has been replicated in subsequent Fall and Winter offerings. Both courses reflect traditional software engineering classroom demographics in Canada, characterized by a majority male student population including both domestic and international students, with smaller but diverse representation from underrepresented groups, including women, Black students, and LGBTQ students. The integration of role models was implemented as an addition to existing technical content without modification to core learning objectives or assessment structures.

### 3.2. Intervention

Role model integration was pre-planned prior to each semester and refined across offerings. Four instructional strategies were employed across multiple lectures and terms. First, selected lectures incorporated brief historical contextualization of technical concepts, highlighting contributors associated with the technologies under study. Second, targeted slides were included to acknowledge pioneers connected to specific technical topics. Third, lectures aligned with institutional awareness periods (e.g., Women in Science, Black History Month, Pride Month) incorporated visibility components featuring contributors from historically underrepresented groups. Fourth, optional readings were provided on influential scientists and engineers in the field. These integrations were not directly tied to assessed learning outcomes and were not included in examinations. Instead, they functioned as contextual enrichment to situate technical knowledge within broader historical and social trajectories.

Peer reviewed

### 3.3. Data Collection

The primary data consist of reflective teaching memos written during the semester and retrospective reflections produced between course offerings to inform adaptation and refinement. Archived lecture slide decks documenting the integration of role model content were consulted as instructional artifacts. Informal student communications, including unsolicited emails and comments referencing inclusivity and belonging, as well as anonymized end-of-term course evaluations, were reviewed as contextual indicators of classroom climate. Classroom observations of student reactions and engagement during lectures were documented in analytic memos. No formal surveys, structured interviews, or direct measures of belonging were conducted. No student data were collected for research purposes; the analysis is grounded exclusively in the instructor's documented experience.

### 3.4. Data Analysis

Analysis followed an iterative interpretive process consistent with analytic autoethnography [2]. First, instructional artifacts and reflective memos were organized chronologically across course offerings beginning in Winter 2024. Second, memos were revisited and subjected to descriptive and open coding to identify recurring patterns related to mechanisms of integration, instructor motivations, perceived student engagement, and indicators of inclusivity and belonging. Third, codes were grouped into higher-level categories representing pedagogical mechanisms and perceived cultural effects within the classroom. Throughout analysis, emerging interpretations were examined in relation to existing literature on role models and belonging in engineering education. The objective was not to establish causal claims but to characterize how embedding role models within technical discourse may function as a pedagogical mechanism influencing classroom inclusivity and sense of belonging.

### 3.5. Reflexivity

The author occupies complete member researcher status as course instructor and Chair of the departmental Equity, Diversity, and Inclusion committee. The author also belongs to underrepresented groups within software engineering, which informs both motivation and interpretive framing. Reflexive memoing was used to interrogate assumptions about belonging and to avoid over-attribution of student responses to the intervention. The initiative emerged from sustained engagement with research on representation and hidden figures in software history. Although the expectation that visibility may influence belonging informed the intervention, no formal hypotheses were tested.

### 3.6. Threats to Validity

This study is context bound and does not seek statistical generalization, consistent with analytic autoethnographic traditions [5]. The findings derive from instructor authored reflective materials developed within two educational settings, which constrains external validity and limits transferability beyond comparable contexts. The author's insider positionality and reliance on retrospective reflection introduce potential interpretive bias and selective recall [2]. To mitigate these concerns, reflective memoing was conducted iteratively across semesters, allowing patterns to be revisited, contrasted, and refined over time. In addition, reflexive documentation was maintained to make analytic decisions and assumptions explicit. Because no formal student level measures were collected, interpretations regarding belonging and inclusion should be understood as theoretically informed inferences rather than empirically validated outcomes.

### 3.7. Ethical Considerations

This study analyzes the author's pedagogical practice through self generated reflective materials. No student data were collected, analyzed, or systematically examined. Informal student feedback and course evaluations are referenced only at an aggregate level as contextual indicators and are neither quoted nor treated as primary data sources. In accordance with institutional research ethics guidelines, formal ethics approval was not required because the data consist exclusively of the author's reflections on teaching practice rather than human participant data.

## 4. FINDINGS

This section reports how the systematic integration of historically grounded role models was implemented within core software architecture instruction and describes observed classroom dynamics, student responsiveness, and iterative refinements across course offerings.

### 4.1. Implementation of Topic Aligned Pioneer Integration

Role model integration was implemented systematically across course offerings beginning in Winter 2024. In a typical semester comprising 12–14 lectures, role model content appeared in approximately 5–6 lectures. Insertions were brief, typically lasting around two minutes. These integrations took two primary forms: (1) embedded contextualization directly linked to technical content, and (2) short acknowledgments aligned with institutional awareness periods, as shown in Figures 1, 2 and 3. Embedded integrations occurred within the technical flow of the lecture. For example, during the introduction to software architecture, Margaret Hamilton was presented in direct connection to the formalization of software

Peer reviewed

engineering as a discipline. In other lectures, pioneers such as Alan Turing were introduced as a "Did you know?" historical insight linking contemporary architectural practices to foundational contributors. Calendar-based acknowledgments occurred during specific awareness periods (e.g., Black History Month, Women in Science, Pride Awareness Days) and were positioned either in the middle or at the end of lectures when not directly tied to the technical topic. These acknowledgments were concise and did not alter core learning objectives or assessment structures. Across offerings since Winter 2024, approximately 700 students have been exposed to this practice.

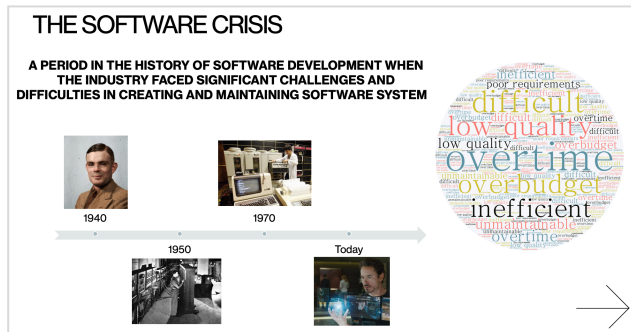

**Figure 1:** Role model incorporation in topic introduction.

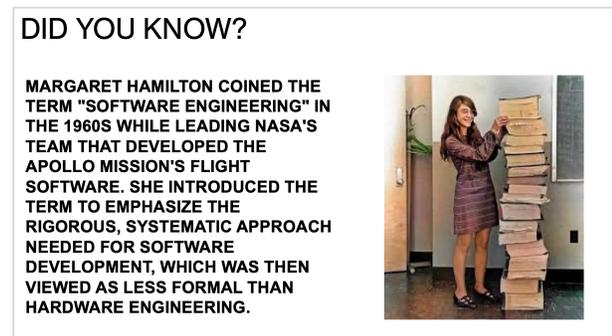

**Figure 2:** "Did you know" strategy for role model introduction.

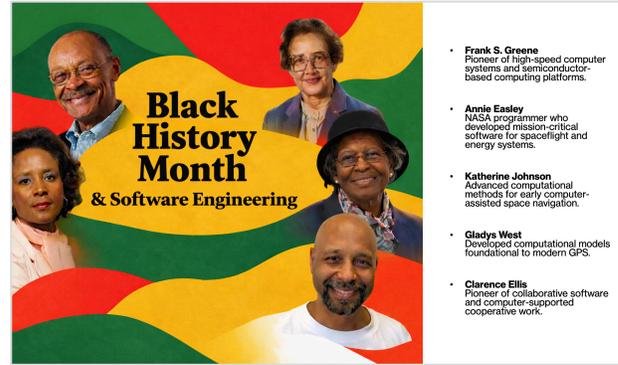

**Figure 3:** Back History month role model introduction.

## 4.2. Pioneers Integrated in the Software Architecture Course

The pioneers included in the Software Architecture course were selected from the research-informed pool identified in prior work on hidden figures in software engineering. The following individuals were incorporated depending on the lecture topic:

- Ada Lovelace (Woman) – Recognized for conceptualizing the first algorithm intended for machine execution, establishing foundational ideas about programmability.
- Alan Turing (Gay man) – Formalized theoretical models of computation that underpin algorithmic reasoning, system design, and artificial intelligence.
- Annie Easley (Black woman) – Developed computational systems for NASA energy and aerospace applications, contributing to large-scale scientific software systems.
- Christopher Strachey (Gay man) – Contributed to programming language theory and early operating systems concepts, informing abstraction and modular reasoning.
- Clarence Ellis (Black man) – Pioneered collaborative software systems and computer-supported cooperative work, relevant to distributed and service-based architectures.
- Danielle Bunten Berry (Transgender woman) – Contributed to human-centered and multiplayer system design, highlighting social interaction within software systems.
- Dennis Ritchie (Not from an underrepresented group) – Co-created the C programming language and contributed to UNIX, shaping system-level programming and operating system architecture.
- Dorothy Vaughan (Black woman) – Contributed to computational transition processes from manual to machine-based systems at NASA, supporting early computing infrastructure.

Peer reviewed

- Frank Greene (Black man) – Advanced high-speed computing systems and semiconductor-based computing platforms, influencing hardware–software integration.
- Grace Hopper (Woman) – Developed early compiler technology and advanced machine-independent programming languages, influencing abstraction layers central to architectural design.
- John von Neumann (Not from an underrepresented group) – Formalized the stored-program computer architecture model, providing the conceptual foundation for modern computer system organization.
- Katherine Johnson (Black woman) – Applied advanced computational methods foundational to aerospace navigation systems and large-scale numerical computation.
- Margaret Hamilton (Woman) – Led development of Apollo flight software and formalized disciplined software engineering practices.
- Peter Landin (Bisexual man) – Advanced programming language semantics and functional paradigms that influence architectural modularity and abstraction.

These figures were not presented as symbolic representations but as contributors whose work shaped foundational aspects of programming, abstraction, software theory, and large-scale systems.

### 4.3. Student Reactions and Classroom Climate

Across offerings since 2024, approximately four to five course evaluations per semester spontaneously referenced inclusivity or classroom climate positively. These comments were not prompted by direct survey questions about diversity or representation. Representative student statements included: 1) Students describing the course as creating a "welcoming and inclusive classroom environment." 2) Comments noting that there was "no pressure in speaking up" and that students "never felt left out." 3) References to the class as an "inclusive environment where everyone could ask questions." In addition to formal evaluations, approximately two to three unsolicited emails per semester referenced appreciation for the inclusive environment or classroom climate. No student questioned the relevance of including historical contributors. Based on classroom observations documented in reflective memos, students appeared attentive during these brief contextual slides. The integration did not generate visible resistance and did not disrupt lecture pacing.

### 4.4. Challenges and Trade-offs

The integration did not generate measurable time pressure, as contextual slides were brief and embedded within existing technical explanations. However, preparation required additional effort to identify relevant pioneers and verify historical contributions to ensure accuracy and relevance. A potential risk of superficiality was considered, particularly during calendar-based acknowledgments not directly tied to lecture content. This concern informed subsequent refinement toward stronger topic alignment in later offerings. Importantly, because the practice was integrated as historical contextualization rather than standalone diversity instruction, no tension was observed between technical rigor and contextual enrichment.

### 4.5. Iterative Refinement Across Offerings

The practice evolved across semesters. In early implementations, role model slides were often positioned at the beginning of lectures. Over time, placement shifted strategically toward moments of conceptual transition or toward the end of lectures during awareness periods. This refinement improved flow and reduced potential perception of interruption. Topic alignment also became more deliberate, ensuring that historical contributors were directly connected to architectural principles, abstraction mechanisms, or system design concepts. Through iterative adjustment, the practice stabilized as a low-disruption, repeatable integration strategy embedded within technical instruction.

## 5. DISCUSSION

Role models have been widely investigated in educational research as influential figures in students' identity formation, academic motivation, and long term achievement [6, 10, 16]. Prior studies indicate that exposure to visible and relatable figures can shape expectancy beliefs, belonging perceptions, and, in some cases, course taking behavior, particularly in STEM domains where representation gaps persist [1, 3]. Our study is consistent with this literature in that students spontaneously referenced inclusivity and psychological comfort, suggesting that contextual signals of representation may contribute to classroom climate even when not explicitly assessed.

From a technical autoethnography perspective, our intervention contributes to this line of work by illustrating how role model exposure may operate through routine instructional practices rather than discrete activities. The integration of historically grounded pioneers within technical content appears to function as a repeated, low intensity signal that situates diverse contributors as part of the normative fabric of the discipline. This form of exposure differs from episodic encounters by accumulating over time, which may support belonging through familiarity and normalization rather than through singular impactful events. While we do not directly measure belonging, students' references to psychological comfort

Peer reviewed

and inclusivity suggest that such embedded signals may influence how students interpret their place within the field. This observation should be interpreted cautiously, as the current study does not establish causal mechanisms, but it provides a basis for further empirical investigation of how embedded representation shapes belonging trajectories in engineering education.

An additional consideration relates to potential differences across student populations. While both undergraduate and MEng cohorts were represented in our teaching context, the reflections analyzed did not indicate systematic variation between these groups. In particular, we did not observe differences suggesting that MEng students, including those from international backgrounds, engaged differently with the material or reported distinct challenges related to exposure to LGBTQ topics. At the same time, the study was not designed to support a comparative analysis across subpopulations. Given prior evidence that cultural and educational backgrounds may influence how inclusivity is perceived, this dimension warrants more targeted investigation in future work.

Our practice also differs from most existing interventions in format and intensity. Much of the prior research focuses on structured exposure formats such as mentoring relationships, guest talks, or experimental presentations [1, 10]. In contrast, our strategy embedded historically grounded pioneers directly within technical explanations of software engineering principles. Rather than introducing separate diversity sessions, our practice integrated brief, topic aligned contextualizations into the flow of core content without modifying learning objectives or assessments. This structural embedding distinguishes our approach from episodic or externally framed role model interventions.

The contribution of this practice lies in demonstrating a low disruption, repeatable integration model within a required software engineering course. Rather than treating representation as an add on, our strategy frames diverse contributors as constitutive of disciplinary history. While the study remains descriptive and does not establish causal effects on motivation or persistence, the observations suggest that historically grounded integration can coexist with technical rigor and may function as a stable contextual signal within software engineering education.

Several lessons emerged from the implementation. First, alignment between the pioneer's contribution and the technical topic supports credibility. Second, brevity supports feasibility, as short integrations reduce time pressure while allowing cumulative exposure. Third, iterative refinement across offerings improves placement and coherence. These elements indicate that other engineering instructors can adopt a similar approach by identifying historically grounded contributors whose work directly shaped core concepts in their field, embedding concise contextual slides within existing lectures, and refining placement based on classroom flow.

## 6. CONCLUSIONS AND FUTURE WORK

This paper presented an analytic autoethnographic account of the systematic integration of historically grounded role models within core software engineering instruction across undergraduate and graduate courses. Drawing on reflective memos and instructional artifacts, we characterized how embedding brief, topic-aligned pioneer contextualizations within routine technical lectures operated as a low-disruption pedagogical practice. By aligning historical contributors directly with architectural principles, abstraction mechanisms, and system design concepts, our practice positioned diverse participation as constitutive of software engineering's development rather than as an external addition. Overall, this study provides preliminary empirical insight into how structurally embedded role model exposure may function as a contextual signal of inclusivity while preserving technical rigor within engineering education.

As future work, we plan to extend this reflective account through systematic empirical studies incorporating validated measures of belonging, expectancy beliefs, and identity alignment. We also aim to explore whether cumulative embedded exposure influences persistence, course selection, or engagement across different student populations. Replicating and adapting this integration model across additional engineering disciplines and institutional contexts may further demonstrate how historically grounded representation strategies interact with local structures, classroom demographics, and broader educational environments.

## Disclosure Statement

Generative artificial intelligence tools were used solely to support grammar checking and spelling correction during manuscript preparation. All content, including the study design, analysis, interpretation, and writing of the manuscript, was created by the authors. No GenAI tools were used in performing the research or generating original content.

Peer reviewed